\documentclass[10pt, twocolumn]{IEEEtran}

\usepackage{amsmath, mathrsfs,amsfonts}
\usepackage{amsfonts}
\usepackage{amssymb}
\usepackage{acronym}
\usepackage{graphicx}

\def\ignore#1{}

\def\bs{\boldsymbol}

\def\ol{\bar}

\def\tl{\tilde}

\def\rn{\mathbb{R}^n}

\def\foral{\textrm{for all} \ }

\newtheorem{assumption}{Assumption}

\newtheorem{dom_face}{Definition}
\newtheorem{expansion}[dom_face]{Definition}

\newtheorem{approx_proj}[dom_face]{Definition}
\newtheorem{submodularity}[dom_face]{Definition}
\newtheorem{config}[dom_face]{Definition}
\newtheorem{codable}[dom_face]{Definition}
\newtheorem{spin-off}[dom_face]{Definition}
\newtheorem{hyper_user}[dom_face]{Definition}

\newtheorem{convergence-thm}{Theorem}

\newtheorem{mod-algorithm}[convergence-thm]{Theorem}
\newtheorem{stepsize_thm}[convergence-thm]{Theorem}

\newtheorem{approx_proj_prop}{Proposition}
\newtheorem{convergence_prop}[approx_proj_prop]{Proposition}
\newtheorem{SFM}[approx_proj_prop]{Proposition}
\newtheorem{rate-splitting}[approx_proj_prop]{Proposition}

\newtheorem{half-space-proj}{Lemma}
\newtheorem{psuedo-nonexp}[half-space-proj]{Lemma}

\newtheorem{expansion_thm}[half-space-proj]{Lemma}

               {\begin{list}{}{\leftmargin#1\rightmargin#2\topsep#3}\item{}}%
               {\end{list}}

\begin{document}
\title{Resource Allocation in Multiple Access Channels\thanks{This research was partially supported by the
National Science Foundation under grant DMI-0545910, and by DARPA
ITMANET program.}}


\author{Ali ParandehGheibi\thanks{A.\ ParandehGheibi is with the Laboratory for
Information and Decision Systems, Electrical Engineering and
Computer Science Department, Massachusetts Institute of Technology,
Cambridge MA, 02139 (e-mail: parandeh@mit.edu)}, Atilla
Eryilmaz\thanks{A.\ Eryilmaz is with the Electrical and Computer
Engineering, Ohio State University, OH, 43210 (e-mail:
eryilmaz@ece.osu.edu)}, Asuman Ozdaglar, and Muriel M\'edard\thanks{
A.\ Ozdaglar and M.\ M\'edard are with the Laboratory for
Information and Decision Systems, Electrical Engineering and
Computer Science Department, Massachusetts Institute of Technology,
Cambridge MA, 02139 (e-mails: asuman@mit.edu, medard@mit.edu)}}


\maketitle \thispagestyle{headings}

\begin{abstract}
We consider the problem of rate allocation in a Gaussian
multiple-access channel, with the goal of maximizing a utility
function over transmission rates. In contrast to the literature
which focuses on linear utility functions, we study general concave
utility functions. We present a gradient projection algorithm for
this problem. Since the constraint set of the problem is described
by exponentially many constraints, methods that use exact
projections are computationally intractable. Therefore, we develop a
new method that uses approximate projections. We use the polymatroid
structure of the capacity region to show that the approximate
projection can be implemented by a recursive algorithm in time
polynomial in the number of users. We further propose another
algorithm for implementing the approximate projections using
rate-splitting and show improved bounds on its convergence time.
\end{abstract}

\section{Introduction}
Dynamic allocation of communication resources such as bandwidth or
transmission power is a central issue in multiple access channels in
view of the time varying nature of the channel and interference
effects. Most of the existing literature on resource allocation in
multiple access channels focuses on specific communication schemes
such as TDMA (time-division multiple access) \cite{TDMA} and CDMA
(code-division multiple access) \cite{CDMA1,CDMA3} systems. An
exception is the work by Tse \emph{et al.} \cite{Tse}, who
introduced the notion of \emph{throughput capacity} for the fading
channel with Channel State Information (CSI) and studied dynamic
rate allocation policies with the goal of maximizing a linear
utility function of rates over the throughput capacity region.

In this paper, we consider the problem of rate allocation in a multiple access channel with perfect
CSI. Contrary to the linear case in \cite{Tse}, we consider maximizing a general utility function
of transmission rates over the capacity region. General concave utility functions allow us to model
different performance metrics and fairness criteria (cf.\ Shenker \cite{She95}, Srikant
\cite{Srikant}). In view of space restrictions, we focus on the non-fading channel in this paper.
In our companion paper \cite{MAC_Wiopt}, we extend our analysis to the fading channel.

Our contributions can be summarized as follows.

We introduce a gradient projection method for the problem of
maximizing a concave utility function of rates over the capacity
region of a non-fading channel. We establish the convergence of the
method to the optimal solution of the problem. Since the capacity
region of the multiple-access channel is described by a number of
constraints exponential in the number of users, the projection
operation used in the method can be computationally expensive. To
reduce the computational complexity, we introduce a new method that
uses \emph{approximate projections}. By exploiting the polymatroid
structure of the capacity region, we show that the approximate
projection operation can be implemented in polynomial time using
submodular function minimization algorithms. Moreover, we present a
more efficient algorithm for the approximate projection problem
which relies on rate-splitting \cite{Urbanke}. This algorithm also
provides the extra information that allows the receiver to decode
the message by successive cancelation.

Other than the papers cited above, our work is also related to the work of Vishwanath \emph{et al.}
\cite{Vishwanath} which builds on \cite{Tse} and takes a similar approach to the resource
allocation problem for linear utility functions. Other works address different criteria for
resource allocation including minimizing the weighted sum of transmission powers \cite{power_min},
and considering Quality of Service (QoS) constraints \cite{QoS}. In contrast to this literature, we
consider the utility maximization framework for general concave utility functions.

The remainder of this paper is organized as follows: In Section II, we introduce the model and
describe the capacity region of a multiple-access channel. In Section III, we consider the utility
maximization problem in non-fading channel and present the gradient projection method. In Section
IV, we address the complexity of the projection problem. Finally, we give our concluding remarks in
Section V.

Regarding the notation, we denote by $x_i$ the $i$-th component of a
vector $\bs x$. We denote the nonnegative orthant by
$\mathbb{R}^n_+$, i.e., $\mathbb{R}^n_+ = \{\bs x\in
\mathbb{R}^n\mid \bs x\ge 0\}$. We write $\bs x'$ to denote the
transpose of a vector $\bs x$. We use $\|x\|$ to denote the standard
Euclidean norm, $\|x\|=\sqrt{x'x}$, and $\mathcal P(\ol x)$ to
denote the exact projection of a vector $\ol x\in \rn$ on a nonempty
closed convex set $X$, i.e.,
\[\mathcal P(\ol x)=\arg\min_{x\in X}\|\ol x-x\|.\]

\section{System Model}
We consider $M$ users sharing the same media to communicate to a single receiver. We model the
channel as a Gaussian multiple access channel with flat fading effects
\begin{equation}\label{fading_model}
    Y(n) = \sum_{i=1}^M  \sqrt{H_i(n)} X_i(n) + Z(n),
\end{equation}
where $X_i(n)$ are the transmitted waveform with average power $P_i$, $H_i(n)$ is the channel gain
corresponding to the \textit{i}-th user and $Z(n)$ is white Gaussian noise with variance $N_0$. We
assume that the channel gains are known to all users and the receiver \footnote{This is assumption
is satisfied in practice when the receiver measures the channels and feeds back the channel
information to the users.}.

We focus on the non-fading case when the channel gains are fixed. We assume without loss of
generality that all channel gains are equal to unity. The capacity region of the Gaussian
multiple-access channel is described as follows \cite{cover}:
\begin{eqnarray}\label{Cg}
    C_g(\bs P) &=& \bigg\{ \bs R \in \mathbb{R}^M_+: \sum_{i \in S} R_i \leq  C\Big(\sum_{i \in S} P_i,
    N_0\Big), \nonumber \\
     &&\qquad \qquad  \textrm{for all}\  S \subseteq \mathcal M = \{1,\ldots, M\} \bigg\},
\end{eqnarray}
where $P_i$ and $R_i$ are the \emph{i}-th user's power and rate,
respectively. $C(P,N)$ denotes Shannon's formula for the capacity of
an AWGN channel given by
\begin{equation}\label{C_AWGN}
    C(P,N) = \frac{1}{2}\log(1+\frac{P}{N}) \quad \textrm{nats}.
\end{equation}

\section{Resource Allocation in Non-fading Channel}
Consider the following utility maximization problem in a $M$-user non-fading multiple-access
channel with channel gains fixed to unity.
\begin{eqnarray}\label{RAC}
   \textrm{maximize} &&u(\bs R)  \nonumber \\
  \textrm{subject to}&& \bs R \in C_g(\bs P),
\end{eqnarray}
where $R_i$ and $P_i$ are $i$-th user rate and power, respectively.
The utility function $u(\bs R)$ is assumed to satisfy the following
conditions.
\begin{assumption}\label{assumption_u} \emph{
\begin{itemize}
\item[(a)] The utility function $u(\bs R)$ is concave with respect to vector $\bs R$.
\item[(b)] The utility function $u(\bs R)$ is monotonically non-decreasing with respect to $R_i$, for $i = 1, \ldots , M$.
\item[(c)] There exists a scalar $B$ such that
$$\|\bs g \| \leq B, \quad \textrm{for all}\ \bs g \in \partial u(\bs R),$$
where $\partial u(\bs R)$ denotes the subdifferential of $u$ at $\bs R$.
\end{itemize}
}\label{utility_assump}
\end{assumption}

The maximization problem in (\ref{RAC}) is a convex program and the
optimal solution can be obtained by several variational methods such
as the gradient projection method. The gradient
    projection method with exact projection is typically user for problems where the projection operation is simple,
    i.e., for problems with simple constraint sets such as
    the non-negative orthant or a simplex. However, the constraint set in (\ref{RAC}) is defined by
    exponentially many constraints, making the projection problem computationally intractable.
    To alleviate this problem, we use an approximate projection, which is  obtained by successively projecting on
    some violated constraint.

\begin{approx_proj}\label{approx_proj_def}
    Let $X = \{\bs x \in \mathbb{R}^n| A\bs x \leq \bs b\}$ where $A$ has non-negative entries. Let $\bs y \in
    \mathbb{R}^n$ violate the constraint $\bs a_i' \bs x \leq b_i$, for $i\in\{i_1, \ldots, i_l\}$. The approximate
    projection of $\bs y$ on $X$, denoted by $\tilde{\mathcal P}$, is given by
    $$ \tilde{\mathcal P}(\bs y) = \mathcal P_{i_1}(\ldots (\mathcal P_{i_{l-1}}(\mathcal P_{i_l}(\bs y)))),$$
    where $\mathcal P_{i_k}$ denotes the exact projection on the hyperplane $\{\bs x \in \mathbb{R}^n| \bs a_{i_k}' \bs x =
    b_{i_k}\}$.
\end{approx_proj}

\begin{approx_proj_prop}\label{approx_proj_prop}
The approximate projection $\tilde{\mathcal P}$ given in Definition
\ref{approx_proj_def} has the following properties:
\begin{itemize}
\item[(i)] For any $\bs y \in \mathbb{R}^n$, $\tilde{\mathcal P}(\bs y)$ is feasible with respect to set $X$, i.e., $\tilde{\mathcal P}(\bs y)\in X$.
\item[(ii)] $\tilde{\mathcal P}$ is pseudo-nonexpansive, i.e.,
\begin{equation}\label{nonexpan}
    \|\tilde{\mathcal P}(\bs y) - \bs{\tl y} \| \leq \|\bs y - \bs{ \tl y}\|,\quad  \textrm{for all} \
\bs {\tl y} \in X.
\end{equation}

\end{itemize}
\end{approx_proj_prop}
\begin{proof}
For part (i), it is straightforward to see that $\mathcal P_i(\bs
y)$ is given by (c.f. \cite{nlp} Sec.\ 2.1.1)
$$\mathcal P_i(\bs y) = \bs y - \frac{\bs a_i'\bs y - b_i}{\|\bs a_i\|}\bs a_i.$$
Since $\bs a_i$ just has non-negative entries, all components of
$\bs y$ are decreased after projection and hence, the constraint $i$
will not be violated in the subsequent projections. Given an
infeasible vector $y\in \mathbb{R}^n$, the approximate projection
operation given in Definition \ref{approx_proj_def} yields a
feasible vector with respect to set $X$.

Part (ii) can be verified using the nonexpansiveness property of projection on a convex set. Since
$\bs {\tl y} $ is a fixed point of $\mathcal P_i$ for all $i$, we have
        \begin{eqnarray}
           \|\tilde{\mathcal P}(\bs y) - \bs {\tl y}\| &=&  \|\mathcal P_{i_1}(\ldots (\mathcal P_{i_l}(\bs y))) - \mathcal P_{i_1}(\ldots (\mathcal P_{i_l}(\bs {\tl y})))\| \nonumber \\
           &\leq& \|\mathcal P_{i_2}(\ldots (\mathcal P_{i_l}(\bs y))) - \mathcal P_{i_{2}}(\ldots (\mathcal P_{i_l}(\bs {\tl y})))\| \nonumber \\
           & \vdots & \nonumber \\
           &\leq& \|\bs y-\bs {\tl y}\|.
        \end{eqnarray}

\end{proof}

     Note that the result of approximate projection depends on the order of projections on violated
     constraints and hence it is not unique. The $k$-th iteration of the gradient projection method
     with approximate projection is given by
    \begin{equation}\label{iteraion}
        \bs R^{k+1} = \tilde{\mathcal P}(\bs R^k + \alpha ^k \bs g^k), \quad \bs g^k \in \partial
        u(\bs R^k),
    \end{equation}
    where $\bs g^k$ is a subgradient at $\bs R^k$, and $\alpha ^k$ denotes
    the stepsize. The following theorem provides a sufficient condition that can be used to
    establish convergence to the optimal solution.

    \begin{convergence-thm}\label{convergence-thm}
        Let Assumption \ref{assumption_u} hold, and $\bs R^*$ be an optimal solution of problem (\ref{RAC}). Also, let the sequence $\{\bs R^k\}$ be
        generated by the iteration in (\ref{iteraion}). If the stepsize $\alpha^k$ satisfies
        \begin{equation}\label{stepsize}
            0 < \alpha^k < \frac{2\left(u(\bs R^*) - u(\bs R^k)\right)}{\|\bs g^k\|^2},
        \end{equation}
then
        \begin{equation}\label{contraction}
            \|\bs R^{k+1} - \bs R^*\| < \|\bs R^k - \bs R^*\|.
        \end{equation}
    \end{convergence-thm}

    \begin{proof}
        We have
        \begin{eqnarray}
            \|\bs R^k + \alpha ^k \bs g^k - \bs R^*\|^2 &=& \|\bs R^k  - \bs R^*\|^2  + 2 \alpha^k (\bs R^k - \bs R^*)' \bs g^k \nonumber \\
            && + (\alpha^k)^2 \| \bs g^k\|^2. \nonumber
        \end{eqnarray}
        By concavity of $u$, we have
        \begin{equation}\label{conv2}
            (\bs R^* - \bs R^k)' \bs g^k \geq u(\bs R^*) - u(\bs R^k).
        \end{equation}
        Hence,
        \begin{eqnarray}
            \|\bs R^k + \alpha ^k \bs g^k - \bs R^*\|^2 &\leq& \|\bs R^k  - \bs R^*\|^2   \nonumber \\
            && \!\!\!\!\!\!\!\!\!\!\!\!\!\!\!\! - \alpha^k \left[ 2\left(u(\bs R^*) - u(\bs R^k)\right)- (\alpha^k) \| \bs g^k\|^2\right]. \nonumber
        \end{eqnarray}
    If the stepsize satisfies (\ref{stepsize}), the above relation yields the following
        $$ \|\bs R^k + \alpha ^k \bs g^k - \bs R^*\| < \|\bs R^k  - \bs R^*\|. $$
        Now by applying pseudo-nonexpansiveness of the approximate projection we have
        \begin{eqnarray}\label{conv3}
            \|\bs R^{k+1} - \bs R^*\| &=& \| \tilde{\mathcal P}(\bs R^k + \alpha ^k \bs g^k) - \bs R^* \| \nonumber \\
            && \leq \|\bs R^k + \alpha ^k \bs g^k - \bs R^*\| < \|\bs R^k  - \bs R^*\|. \nonumber
        \end{eqnarray}
    \end{proof}
\begin{convergence_prop}\label{convergence_prop}
Let Assumption \ref{assumption_u} hold. Also, let the sequence
$\{\bs R^k\}$ be
        generated by the iteration in (\ref{iteraion}). If the stepsize $\alpha^k$ satisfies
        (\ref{stepsize}),
then $\{\bs R^k\}$ converges to an optimal solution $\bs R^*$.
\end{convergence_prop}
\begin{proof}
See Proposition 8.2.7 of \cite{convexbook}.
\end{proof}

The convergence analysis for this method can be extended for
different stepsize rules. For instance, we can employ diminishing
stepsize, i.e., $$ \alpha^k \rightarrow 0, \qquad \sum_{k =
0}^{\infty} \alpha^k = \infty,$$ or more complicated dynamic
stepsize selection rules such as the \textit{path-based incremental
target level} algorithm proposed by Br\"{a}nnlund \cite{brann} which
guarantees convergence to the optimal solution, and has better
convergence rate compared to the diminishing stepsize rule.

\section{Complexity of the Projection Problem}
    Even though the approximate projection is simply obtained by successive projection on the
    violated constraints, it requires to find the violated constraints among exponentially many
    constraints describing the constraint set. In this section, we exploit the special structure
    of the capacity constraints so that each gradient projection step in (\ref{iteraion}) can be performed in
    polynomial time in $M$.

\begin{submodularity}\label{submodularity}
Let $f:2^\mathcal M \rightarrow \mathbb R$ be a function defined over all subsets of $\mathcal M$.
$f$ is \emph{submodular}  if
    \begin{equation}\label{submodular_def}
        f(S \cup T) + f(S \cap T) \leq f(S) + f(T), \quad \textrm{for all}\  S,T \in 2^\mathcal{M}.
    \end{equation}
\end{submodularity}

\begin{SFM}\label{SFM_prop}
For any $\bs{\bar{R}} \in \mathbb{R}_+^M$, finding the most violated capacity constraint in
(\ref{Cg}) is equivalent to a \emph{submodular function minimization} (SFM) problem.
\end{SFM}
\begin{proof}
Define $f_C(S): 2^\mathcal M \rightarrow \mathbb R$ as follows
\begin{equation}\label{fC}
f_C(S) = C(\sum_{i \in S} P_i, N_0), \quad \foral S \subseteq \mathcal M.
\end{equation}

 It is straightforward to see that
$f_C$ is a submodular function.

We can rewrite the capacity constraints in (\ref{Cg}) as
    \begin{equation}\label{capacity_region2}
         f_C(S) - \sum_{i \in S} R_i \geq 0, \quad \foral S
        \subseteq \mathcal{M}.
    \end{equation}
Thus, the most violated constraint at $\bs{\bar{R}}$ is given by
    \begin{eqnarray}\label{SFM}
        S^* = \rm arg \!\min_{S \in 2^\mathcal{M}} &&  f_C(S) -  \sum_{i \in S} \overline{R}_i. \nonumber
    \end{eqnarray}
Since summation of a submodular and a linear function is also submodular, the problem above is of
the form of submodular function minimization.
\end{proof}

    It is first shown by Gr\"{o}tschel \emph{et al.} \cite{Grotschel}  that SFM problem can be solved in
    strongly polynomial time. The are several fully combinatorial strongly polynomial algorithms in the literature.
    The best known algorithm for SFM proposed by Orlin \cite{orlin} has running time $O(M^6)$ for the submodular function defined in (\ref{fC}). Note
    that approximate projection does not require any specific order for successive projections.
    Hence, finding the most violated constraint is not necessary for approximate projection. In
    view of this fact, a more efficient algorithm based on rate-splitting is presented in Appendix \ref{appendix}, to find
    a violated constraint. This algorithm runs in $O(M^2 \log M)$ time.

    Although a violated constraint can be obtained in polynomial time, it does not guarantee that
    the approximate projection can be performed in polynomial time. Because it is possible to have
    exponentially many constraints violated at some point and hence the total running time of the
    projection would be exponential in $M$. However, we show that for small enough stepsize in the gradient
    projection iteration (\ref{iteraion}), no more than $M$ constraints can be violated at each
    iteration. Let us first define the notion of expansion for a polyhedron.

        \begin{expansion}\label{expansion_def}
        Let $Q$ be a polyhedron described by a set of linear constraints, i.e.,
        \begin{equation}\label{polyhedron}
            Q = \left\{\bs x \in \mathbb{R}^n: A \bs x \leq \bs b \right\}.
        \end{equation}
        Define the \emph{expansion} of $Q$ by $\delta$, denoted by $\mathcal{E}_\delta(Q)$, as the polyhedron
        obtained by relaxing all the constraints in (\ref{polyhedron}), i.e., $ \mathcal{E}_\delta(Q) = \left\{\bs x \in \mathbb{R}^n: A \bs x \leq \bs b + \delta\mathbf{1}
            \right\},$
        where $\mathbf{1}$ is the vector of all ones.
\end{expansion}

    \begin{expansion_thm}\label{expansion_thm}
        Let $f_C$ be as defined in (\ref{fC}). There exists a positive scalar $\delta$ satisfying
            \begin{eqnarray}\label{exp_lemma_hyp}
                \delta &\leq& \frac{1}{2} (f_C(S)+ f_C(T) - f_C(S \cap T) - f_C(S \cup T)), \nonumber \\
                && \quad \quad \quad \foral S,T \in 2^\mathcal{M}, \quad   S   \cap T \neq
                S,T,
            \end{eqnarray}
such that any point in the relaxed capacity region of an $M$-user
multiple-access channel,
        $\mathcal{E}_{\delta}(C_g)$, violates no more than $M$ constraints of $C_g$ defined in (\ref{Cg}).
    \end{expansion_thm}

    \begin{proof}
        Existence of a positive scalar $\delta$ satisfying (\ref{exp_lemma_hyp}) follows from
        submodularity  of $f_C$, and the fact that neither $S$ nor $T$ contains the other one.

        Suppose for some $\bs R \in \mathcal{E}_{\delta}(C_g)$, there are $M+1$ constraints of
        $C_g$ violated. There are at least two violated constraints corresponding to some
        sets $S,T \in 2^ \mathcal M$ where $S \cap T \neq S,T$. Because it is not possible to have $M+1$
        non-empty nested sets in $2^\mathcal{M}$. We have
        \begin{eqnarray}
          -\sum_{i \in S} R_i &<& -f_C(S), \\
          -\sum_{i \in T} R_i &<& -f_C(T).
        \end{eqnarray}
        Since $\bs R$ is feasible in the relaxed region,
        \begin{eqnarray}\label{cap_const}
          \sum_{i \in S \cap T} R_i &\leq& f_C(S \cap T) + \delta, \\
          \sum_{i \in S \cup T} R_i &\leq& f_C(S \cup T) + \delta.
        \end{eqnarray}
            Note that if $S \cap T = \emptyset$, (\ref{cap_const}) reduces to $0 \leq \delta$ which is
            a valid inequality.

            By summing the above inequalities we conclude
            \begin{equation}\label{contrad_lemma}
                \delta > \frac{1}{2} (f_C(S)+ f_C(T) - f_C(S \cap T) - f_C(S \cup T)),
            \end{equation}
            which is a contradiction.
    \end{proof}

    \begin{stepsize_thm}\label{convtime}
Let Assumption \ref{assumption_u} hold. Let $P_1 \leq P_2 \leq
\ldots \leq P_M$ be the transmission powers.

        If the stepsize $\alpha^k $ in the $k$-th iteration (\ref{iteraion}) satisfies
            \begin{eqnarray}\label{step_lemma_hyp}
                \alpha^k &\leq& \frac{1}{4 B \sqrt{M}} \log \Big[1+\frac{P_1 P_2}{(N_0+\sum_{i=3}^M P_i) (N_0+\sum_{i=1}^M P_i)}
            \Big], \nonumber
            \end{eqnarray}
        then at most $M$ constraints of the capacity region $C_g$ can be violated at each iteration step.
    \end{stepsize_thm}
    \begin{proof} We first show that the inequality in (\ref{exp_lemma_hyp}) holds for the
following choice
        of $\delta$:
        \begin{eqnarray}\label{min_delta}
            && \delta = \frac{1}{4} \log \Big[1+\frac{P_1 P_2}{(N_0+\sum_{i=3}^M P_i) (N_0+\sum_{i=1}^M P_i)}
            \Big]. \nonumber
        \end{eqnarray}
        In order to verify this, rewrite the right hand side of (\ref{exp_lemma_hyp}) as
        \begin{eqnarray}
          && \frac{1}{4} \log \Big[\frac{(N_0+\sum_{i \in S} P_i) (N_0+\sum_{i \in T} P_i)}{(N_0+\sum_{i \in S \cap T} P_i) (N_0+\sum_{i \in S \cup T} P_i)} \Big]  \nonumber \\
           && = \frac{1}{4} \log \Big[1+\frac{\sum_{(i,j) \in (S\setminus T)\times (T \setminus S)} P_i P_j}{(N_0+\sum_{i \in S \cap T} P_i) (N_0+\sum_{i \in S \cup T} P_i)} \Big] \nonumber \\
           && \geq \frac{1}{4} \log \Big[1+\frac{P_1 P_2}{(N_0+\sum_{i \in S \cap T} P_i) (N_0+\sum_{i \in S \cup T} P_i)} \Big] \nonumber \\
           && \geq \frac{1}{4} \log \Big[1+\frac{P_1 P_2}{(N_0+\sum_{i \in S \cap T} P_i) (N_0+\sum_{i=1}^M P_i)} \Big] \nonumber \\
           && \geq \frac{1}{4} \log \Big[1+\frac{P_1 P_2}{(N_0+\sum_{i=3}^M P_i) (N_0+\sum_{i=1}^M P_i)} \Big]. \nonumber
        \end{eqnarray}

        The inequalities can be justified by using the monotonicity of the logarithm function and the fact
        that $(S\setminus T)\times (T \setminus S)$ is non-empty because $S \cap T \neq S,T$.

        Now, let $\bs R^k$ be feasible in the capacity region, $C_g$. For every $S \subseteq \mathcal M$, we have
        \begin{eqnarray}\label{stp_thm_1}
          \sum_{i \in S} (R^k_i + \alpha^k g^k_i) &=& \sum_{i \in S} R^{k}_i + \alpha^k \|g^{k}\|  \sum_{i \in S} \frac{g^{k}_i}{\|g^{k}\| } \nonumber \\
          &\leq& f(S) + \frac{\delta}{B \sqrt{M}} B \sum_{i \in S} \frac{g^{k}_i}{\|g^{k}\| }
          \nonumber \\
          &\leq& f(S) + \delta,
        \end{eqnarray}
        where the first inequality follows from the hypotheses and the second inequality follows from the fact that for any unit vector  $\bs d \in \mathbb R^M$,
        it is true that
        \begin{equation}\label{stp_thm_2}
            \sum_{i \in S} d_i \leq \sum_{i \in S} |d_i| \leq \sqrt{M}.
        \end{equation}
        Thus, if $\alpha^k$ satisfies (\ref{step_lemma_hyp}) then $(\bs R^k + \alpha^k \bs g^k) \in
        \mathcal{E}_{\delta}(C_g)$, for some $\delta$ for which (\ref{exp_lemma_hyp}) holds.
        Therefore, by Lemma \ref{expansion_thm} the number of violated
        constraints does not exceed $M$.
        \end{proof}
In view of the fact that a violated constraint can be identified in $O(M^2\log M)$ time (see the
Algorithm in Appendix I), Theorem \ref{convtime} implies that, for small enough stepsize, the
approximate projection can be implemented in $O(M^3\log M)$ time.

\section{Conclusion}
We addressed the problem of optimal rate allocation in a non-fading
multiple access channel from an information theoretic point of view.
We formulated the problem as maximizing a general concave utility
function of transmission rates over the capacity region of the
multiple-access channel.

We presented an iterative gradient projection method for solving
this problem. In order to make the projection on a set defined by
exponentially many constraints tractable, we considered a method
that uses approximate projections. Using the special structure of
the capacity region, we showed that the approximate projection can
be performed in time polynomial in the number of users.

In ongoing work, we extend our analysis to finding dynamic resource
allocation policies in fading multiple-access channels. We study
both rate and power allocation policies under different assumptions
on the availability of channel statistics information.


\appendices
\section{Algorithm for finding a violated constraint }\label{appendix}

In this section, we present an alternative algorithm based on
rate-splitting idea to identify a violated constraint for an
infeasible point. For a feasible point, the algorithm provides
information for decoding by successive cancellation. We first
introduce some definitions.

\begin{config}\label{config}
   The quadruple $(M, \bs P, \bs R, N_0)$ is called a \textit{configuration} for an $M$-user
    multiple-access channel, where $\bs R = (R_1, \ldots, R_M)$ is the rate tuple, $\bs P = (P_1, \ldots, P_M)$ represents the
    received power and $N_0$ is the noise variance. For any given configuration, the \emph{elevation}, $\bs \delta \in \mathbb R^M$, is
    defined as the unique vector satisfying
    \begin{equation}\label{elevation}
                R_i = C(P_i, N_0+ \delta_i), \quad i =1,\ldots, M.
    \end{equation}
\end{config}
    Intuitively, we can think of message $i$ as rectangles of height $P_i$, raised above the noise
    level by $\delta_i$. In face, $\delta_i$ is the amount of additional Gaussian interference that
    message $i$ can tolerate.
    Note that if a configuration is feasible then its elevation vector is non-negative, but that is
    not sufficient to check feasibility.

\begin{codable}\label{codable}
    The configuration $(M, \bs P, \bs R, N_0)$ is \emph{single-user codable}, if after possible re-indexing,
    \begin{equation}\label{su_codable}
        \delta_{i+1} \geq \delta_i + P_i,   i = 0, 1 , \dots, M-1,
    \end{equation}
    where we have defined $\delta_0 = P_0 = 0$ for convention.
\end{codable}
    By the graphical representation described earlier, a configuration is single-user codable if
    the none of the messages are overlapping.
\begin{spin-off}
    The quadruple $(m, \bs p, \bs r, N_0)$ is a \textit{spin-off} of $(M, \bs P, \bs R, N_0)$ if there exists a
    surjective mapping $\phi : \{1,\ldots, m\} \rightarrow \{1,\ldots, M\}$ such that for all $i \in \{1,\ldots,
    M\}$ we have

\begin{eqnarray}
   P_i &\geq& \sum_{j \in \phi^{-1}(i)} p_j, \nonumber \\
   R_i &\leq& \sum_{j \in \phi^{-1}(i)} r_j. \nonumber
 \end{eqnarray}
    where $\phi^{-1}(i)$ is the set of all $j \in \{1,\ldots,m\}$ that map into $i$ by means of $\phi$.
\end{spin-off}

\begin{hyper_user}\label{hyper_user}
       A \textit{hyper-user} with power $\bar{P}$, rate $\bar{R}$, is obtained by merging $d$ actual users
        with powers $(P_{i_1}, \ldots, P_{i_d})$ and rates $(R_{i_1}, \ldots, R_{i_d})$, i.e,
\begin{equation}\label{hyper}
   \bar{P} = \sum_{k=1}^d P_{i_k}, \quad
          \bar{R} =\sum_{k=1}^d R_{i_k}.
\end{equation}

\end{hyper_user}

\begin{rate-splitting}\label{rate_splitting_thm}
For any $M$-user achievable configuration $(M, \bs P, \bs R, N_0)$, there exists a spin-off $(m,
\bs p , \bs r, N_0)$ which is single user codable.
\end{rate-splitting}
\begin{proof}
See Theorem 1 of \cite{Urbanke}.
\end{proof}
Here, we give a brief sketch of the proof to give intuition about the algorithm. The proof is by
induction on $M$. For a given configuration, if none of the messages are overlapping then the
spin-off is trivially equal to the configuration. Otherwise, merge the two overlapping users into a
\emph{hyper-user} of rate and power equal the sum rate and sum power of the overlapping users,
respectively. Now the problem is reduced to rate splitting for $(M-1)$ users. This proof suggests a
recursive algorithm for rate-splitting that gives the actual spin-off for a given configuration.

It follows directly from the proof of Proposition \ref{rate_splitting_thm} that this recursive
algorithm gives a single-user codable spin-off for an achievable configuration. If the
configuration is not achievable, then the algorithm encounters a hyper-user with negative
elevation. At this point the algorithm terminates. Suppose that hyper-user has rate $\bar R$ and
power $\bar P$. Negative elevation is equivalent to the following
$$\bar R > C(\bar P, N_0).$$
Hence, by Definition \ref{hyper_user} we have,
$$\sum_{i \in S} R_i > C(\sum_{i \in S} P_i, N_0).$$
where $S = \{i_1,\ldots,i_d\} \subseteq \mathcal M$. Therefore, a hyper-user with negative
elevation leads us to a violated constraint in the initial configuration.

The complexity of this algorithm can be computed as follows. The algorithm terminates after at most
$M$ recursions. At each recursion, all the elevations are computed in $O(M)$ time and they are
sorted in $O(M\log M)$ time. Once the users are sorted by their elevation it takes $O(M)$ time to
either find two overlapping users or a hyper-user with negative elevation. Hence, the algorithm
runs in $O(M^2\log M)$ time.

\bibliographystyle{unsrt}
\bibliography{MAC}

\end{document}